\documentclass[12pt, letterpaper]{article}
\usepackage{geometry}
\usepackage{amsmath}
\usepackage{graphicx}
\usepackage{float}
\usepackage{subcaption}
\usepackage{amssymb}
\newgeometry{vmargin={18mm}, hmargin={18mm,18mm}}

\providecommand{\keywords}[1]
{
  \small	
  \textbf{\textit{Keywords---}} #1
}
\begin{document}
\title{Bilayer graphene in periodic and quasiperiodic magnetic superlattices}
\author{David J. Fern\'andez C.$^{1}\footnote{david.fernandez@cinvestav.mx}$, O. Pav\'on-Torres$^{1}\footnote{omar.pavon@cinvestav.mx}$}
\date{%
\small{$^{1}$ Physics Department, Cinvestav, POB 14-740, 07000 Mexico City, Mexico\\
}}
\maketitle

\begin{abstract}
Starting from the effective Hamiltonian arising from the tight binding model, we study the behaviour of low-lying excitations for bilayer graphene placed in periodic external magnetic fields by using irreducible second order supersymmetry transformations. The coupled system of equations describing these excitations is reduced to a pair of periodic Schrödinger Hamiltonians intertwined by a second order differential operator. The direct implementation of more general second-order supersymmetry transformations allows to create nonsingular Schrödinger potentials with periodicity defects and bound states embedded in the forbidden bands, which turn out to be associated to quasiperiodic magnetic superlattices. Applications in quantum metamaterials stem from the ability to engineer and control such bound states which could lead to a fast development of the subject in the near future. 
\end{abstract}

\! \! \! \! \keywords{Bilayer graphene, superlattices, second order SUSY QM, Lam\'e equation.}

\section{Introduction}

The correspondence between band structure of graphene and the topological features of electronic states had played a major role in modern physics since its discovery in 2004 \cite{rev0}. Due to its honeycomb lattice structure, small local changes lead to gauge potentials, associated with the phase of the electronic wavefunction in each of the two independent sublattices \cite{rev1}. In particular, the local in-plane deformations generated by strain can be described by a vector potential, which is equivalent to a pseudomagnetic field applied to graphene. The band structure of graphene can be fundamentally changed in different ways, one of which is the stacking of two layers in order to form bilayer graphene \cite{rev2}. Bilayer graphene presents some advantages over monolayer graphene, since the larger possibilities for tuning experimentally its physical properties \cite{rev3, rev4}. In bilayer graphene there are four atoms per unit cell, with inequivalent sites $A1$, $B1$ and $A2$, $B2$ in the first and second graphene layers, respectively \cite{mik}. The atomic orientation among the two layers might further vary, as bilayer graphene has a weak van der Waals interlayer bonding due to lattice deformation, which largely affects the interlayer electron motion \cite{yang}. Bilayer graphene can display a parabolic dispersion relation at the $K$ points, making electrons behave differently as compared with the single-layer case. Bilayer graphene offers as well the possibility of applying a bias voltage $W$ between the two layers, allowing to tune the band structure \cite{wang}. In particular, the inequivalence of the two graphene layers gives rise to a Mexican-hat-like structure featuring a band gap \cite{rev2}. Let us stress that a tunable gap is important for possible electronic devices.

Different stackings can occur in bilayer graphene. Due to its large stability in bulk graphite, the most studied is the AB Bernal stacking, in which the two graphene layers are arranged in such a way that the $A1$ sublattice is exactly on top of the $B2$ sublattice. On the other hand, in the simple hexagonal or AA stacking, both sublattices of the first layer, $A1$ and $B1$, are located directly on top of the two sublattices $A2$ and $B2$ of the second layer. Although graphene with direct or AA stacking has not been observed in natural graphite, it has been produced by folding graphite layers at the edges of a cleaved sample with a scanning tunnelling microscope tip. A third simple bilayer structure arises when the two layers are stacked with a specific twisting angle ($\theta$), called twisted bilayer graphene (TBG). In this structure the moir\'e patterns with a higher periodicity emerge. The emergence of these patterns features remarkable optical and electronic properties \cite{rev7, rev8, rev9, rev10, rev11}. The electronic structure of TBG shows a linear band dispersion near the Dirac points, rather than the massive quadratic dispersion of the AB-stacked bilayer graphene, suggesting a relatively weak interlayer interaction. In strong magnetic fields, however, it is predicted that the spectrum exhibits a fractal structure, called Hofstadter's butterfly, in which a series of energy gaps appear in a self-similar fashion \cite{butter}. 

Due to its Dirac-like low lying excitations, graphene displays a number of unusual transport properties, which results in the pseudospin 1/2 of the low-lying modes, their linear dispersion relation, and the vanishing density of states at the Dirac points. In contrast, the interlayer interaction in bilayer graphene with regular AB stacking changes the linear dispersion of monolayer graphene into a quadratic one, where an electron behaves as a massive particle \cite{geim}.  In order to describe the situation when magnetic (or pseudomagnetic) fields are applied to monolayer graphene, the first-order supersymmetric quantum mechanics (SUSY QM) is the simplest solution approach \cite{kuru, milpas, MF14,  dav1, dav2, celeita1, ad00, ad1, ad2} (see also \cite{ad01, ad02, ad03, ad04, ad05, ad06, ad07}). On the other hand, the second-order SUSY QM has proved useful to study the charge carriers behaviour of bilayer graphene in external magnetic fields \cite{fer1, fer2}. In addition, the equivalence between Maxwell’s and Dirac equation has been used to understand the electromagnetic spin and orbital angular momentum, and examine the relationship between interface states and topology \cite{ap1, ap2, ap3, andreev2, dom1, dom2}. 

Recently, there has been a growing interest in studying and characterizing transport properties in superlattices composed of stacked van der Waals heterostructures, largely motivated by magic-angle twisted graphene \cite{rv1, rv2, z0, z1, z11, z2, z3, z31, z32, z4}. In this direction, numerous experimental investigations have been realized on dual-gated hexagonal boron nitride and bilayer graphene superlattices, revealing first-order transitions between two insulating states under a perpendicular magnetic field \cite{rv3, rv4, rv5}. On the theoretical side, first-order SUSY QM has been used to study the electronic behavior in superlattices featuring monolayer graphene subjected to periodic magnetic fields \cite{cel1}. It has been observed that the arising of two opposite Darboux displacements disrupts the superlattice's periodicity, although asymptotically the periodicity is recovered. Moreover, the use of such displacements turns out to be fundamental for understanding phenomena related to defects in superlattices.

\medskip

In the present work we aim to study the behaviour of low-lying excitations of bilayer graphene placed in a periodic magnetic field, and its physical implications, by using second order supersymmetric quantum mechanics. We will explore as well more general second-order supersymmetry transformations producing periodicity defects, which will be associated naturally to quasiperiodic magnetic superlatticces for bilayer graphene. In order to do that, the second section of this paper is devoted to some general considerations about the tight-binding model for bilayer graphene, its physical assumptions and the two coupled equations we will arrive for our system. Then, in section 3 a brief overview of the second order SUSY QM will be given, as well as the construction of the intertwining operators when a periodic potential is present. The way to implement the method for generating quasiperiodic partner potentials and the corresponding superpotential will be also discussed. In section 4 we will generate the quasiperiodic magnetic superlattices associated to a physically meaningful quantum problem. Finally, in section 5 we will present our conclusions and the perspectives of this work. 

\section{Tight-binding model for bilayer graphene}

As it was already mentioned, among all possible configurations that bilayer graphene may adopt (AA, AB or twisted bilayer), the AB configuration or Bernal stacking is the most stable one, and it is easily produced in labs. In this configuration, half of the carbon atoms of the top layer are aligned vertically with half of the atoms of the lower layer, while the other carbon atoms of the top layer are located above the centres of the lower-layer hexagons. This means that one layer is rotated with respect to the other by an angle of $\pi/3$, and from now on we will address this configuration only.

\medskip

The effective Hamiltonian around a Dirac point for bilayer graphene is
\begin{equation}
H=\frac{1}{2m^{*}}\begin{pmatrix}
0 & (p_{x}-ip_{y})^{2}\\
(p_{x}+ip_{y})^{2} & 0
\end{pmatrix},
\end{equation}
\noindent where $m^{*}=\gamma_{1}/2v_{F}^{2}\approx 0.054 m_{e}$ is the electron effective mass, $m_{e}$ is the electron free mass, $v_{F}$ is the Fermi velocity,  $p_{x}=-i\hbar\frac{\partial}{\partial x}$ and $p_{y}=-i\hbar\frac{\partial}{\partial y}$. We use the minimal coupling rule to incorporate the vector potential, i.e., $p_{i} \to p_{i}+\dfrac{e}{c}A_{i}$. In the Landau gauge the vector potential can be chosen as $\vec{A}=A(x)\hat{e}_{y}$, implying that $\vec{B}=B(x)\hat{e}_{z}$ thus the magnetic field amplitude takes the form $B(x)=A'(x)$. The eigenvalue equation when an external magnetic field is applied is given by:
\begin{equation}
H \Psi (x, y)=\frac{1}{2m^{*}}\begin{pmatrix}
0 & \Pi^{2}\\
\left(\Pi^{+}\right)^{2} & 0
\end{pmatrix}\Psi(\textbf{x})=E\Psi(\textbf{x}),
\end{equation}
with $\Pi=p_{x}-ip_{y}-i\frac{e}{c}A(x)$. Taking into account the translation invariance along $y$ direction we propose

\begin{equation}
\Psi (x, y)=e^{iky}
\begin{pmatrix}
\psi^{(2)}(x)\\
\psi^{(0)}(x)
\end{pmatrix}. 
\end{equation}
By sticking to the approach introduced in \cite{fer1, fer2}, the following system of equations is obtained:
\begin{equation}
L^{-}_{2}\psi^{(0)}(x)=\left(\frac{d^2}{dx^2}+\eta(x)\frac{d}{dx}+\gamma(x)\right)\psi^{(0)}(x)=-\tilde{E}\psi^{(2)}(x), \label{eli1}
\end{equation}
\begin{equation}
L_{2}^{+}\psi^{(2)}(x)=\left(\frac{d^2}{dx^2}-\eta(x)\frac{d}{dx}+\gamma(x)-\eta^{'}(x)\right)\psi^{(2)}(x)=-\tilde{E}\psi^{(0)}(x), \label{eli2}
\end{equation}
where $\tilde{E}$ and $\eta(x)$ are given by
\begin{equation}
\tilde{E}=\frac{2m^{*}E}{\hbar^{2}}, \qquad \eta(x)=2\left(k+\frac{e}{c\hbar}A(x)\right). \label{eli3}
\end{equation}
Thus, the magnetic field amplitude $B(x)$ takes the form

\begin{equation}
B(x)=\frac{c\hbar}{2e}\eta'(x). \label{kim2}
\end{equation} 
In general, after generating the vector potential $A(x)$ through a second-order SUSY transformation, the amplitude of the magnetic field $B(x)$ (or pseudo-magnetic field) given by expression (\ref{kim2}) can be directly obtained. Depending on the specific physical context under investigation, different forms of $A(x)$ can be proposed, each associated with the superpotential as it will be shown in the following section. Since we want to study the case when a periodic magnetic field is applied, $\eta(x)$ and $\gamma(x)$ in equations (\ref{eli1}) and (\ref{eli2}) are supposed to be complex valued piecewise continuous periodic functions, both with the same period. Note that an iterative first order SUSY treatment could be used, such that the second order differential intertwining operators (\ref{eli1}) and (\ref{eli2}) will be factorized as a product of two in general different first order intertwining operators. These assumptions require to modify the effective Hamiltonian by an additional term, that could be physically interpreted as the result of a trigonal warping effect or a varying external potential. This is expressed mathematically as follows
\begin{equation}
H =\frac{1}{2m^{*}}\begin{pmatrix}
0 & \Pi^{2}\\
\left(\Pi^{+}\right)^{2} & 0
\end{pmatrix}-\dfrac{\hbar^{2}}{2m^{*}}f(x)\sigma_{x},
\end{equation}
where 
\begin{equation}
f(x)=\dfrac{\eta '(x)}{4\eta^{2}(x)}-\dfrac{\eta ''(x)}{2\eta(x)}-\dfrac{(\epsilon_{1}-\epsilon_{2})^{2}}{4\eta^ {2}(x)}.
\end{equation}
The coupled system of equations (\ref{eli1}) and (\ref{eli2}) can be decoupled as follows: 
\begin{equation}
L_{2}^{+}L_{2}^{-}\psi^{(0)}(x)=\tilde{E}^ {2}\psi^{(0)}(x),
\end{equation}
\begin{equation}
L_{2}^{-}L_{2}^{+}\psi^{(2)}(x)=\tilde{E}^ {2}\psi^{(2)}(x).
\end{equation}
We have chosen the index notation slightly different from \cite{fer1, fer2}, for reasons that will be evident below. As can be seen, in order to study the effect of periodic magnetic fields on bilayer graphene, the second order supersymmetric quantum mechanics can be used, as when studying the charge carriers behaviour of bilayer graphene in non-periodic external magnetic fields. Before doing that, however, a brief overview of the second order supersymmetric quantum mechanics will be given in the following section.

\section{Second order supersymmetric quantum mechanics}
Now, we present the algorithm of second order supersymmetric quantum mechanics for showing its explicit relation with bilayer graphene placed in periodic magnetic fields \cite{barbaba}. First, let us assume that $v_{1}(x)$ and $v_{2}(x)$ are two solutions of the stationary Schrödinger equation for the Hamiltonian $H_{0}$ associated to the factorization energies $\epsilon_{1}$ and $\epsilon_{2}$, respectively, \textit{i.e.,} $H_{0}v_{i}=\epsilon_{i} v_{i}$, $i=1, 2$, where
\begin{equation}
H_{0}=-\dfrac{d^ {2}}{dx^{2}}+V_{0}(x) \label{eo1}.
\end{equation}
We will suppose that an intertwining relation involving the two Hamiltonians $H_{0}$, $H_{2}$ and the operator $L_{2}^ {-}$ of equation (\ref{eli1}) is fulfilled, namely, 

\begin{equation}
{H}_{2}L_{2}^{-}=L_{2}^{-}H_{0},
\end{equation}  
where
\begin{equation}
{H}_{2}=-\dfrac{d^ {2}}{dx^{2}}+{V}_{2}(x). \label{eo2}
\end{equation}
We will assume as well that both Hamiltonians (\ref{eo1}) and (\ref{eo2}) have discrete spectra. The so-called SUSY partner potentials $V_{0}(x)$ and $V_{2}(x)$ can be expressed in terms of the functions $\eta(x)$ and $\gamma (x)$ characterizing the interwining operator $L_{2}^ {-}$. After some work, in the first place it is obtained that
\begin{equation}
\gamma (x)=-\dfrac{\eta^{''}(x)}{2\eta(x)}+\left(\dfrac{\eta^{'}(x)}{2\eta (x)}\right)^{2}+\dfrac{\eta^{'}(x)}{2}+\dfrac{\eta^{2}(x)}{4}-\left(\dfrac{\epsilon_{1}-\epsilon_{2}}{2\eta(x)}\right)^{2},
\end{equation}
where $\eta(x)$ is given by
\begin{equation}
\eta(x)=-\dfrac{d}{dx}\ln W(v_{1}(x), v_{2}(x))=\dfrac{(\epsilon_{1}-\epsilon_{2})v_{1}(x)v_{2}(x)}{W(v_{1}(x), v_{2}(x))}, \label{magpot}
\end{equation}
with $W(v_{1}(x), v_{2}(x))\equiv W_{12}(x)$ being the Wronskian of $v_{1}(x)$ and $v_{2}(x)$. The partner potentials $V_{0}(x)$ and ${V}_{2}(x)$ turn out to be expressed as:

\begin{equation}
V_{0}(x)=\dfrac{\eta^{''}(x)}{2\eta(x)}-\left(\dfrac{\eta^{'}(x)}{2\eta (x)}\right)^{2}-\eta^ {'}(x)+\dfrac{\eta^{2}(x)}{4}+\left(\dfrac{\epsilon_{1}-\epsilon_{2}}{2\eta(x)}\right)^{2}+\dfrac{\epsilon_{1}+\epsilon_{2}}{2}, \label{q1}
\end{equation} 

\begin{equation}
V_{2}(x)=\dfrac{\eta^{''}(x)}{2\eta(x)}-\left(\dfrac{\eta^{'}(x)}{2\eta (x)}\right)^{2}+\eta^ {'}(x)+\dfrac{\eta^{2}(x)}{4}+\left(\dfrac{\epsilon_{1}-\epsilon_{2}}{2\eta(x)}\right)^{2}+\dfrac{\epsilon_{1}+\epsilon_{2}}{2}.\label{q2}
\end{equation} 
Notice that the new potential $V_{2}(x)$ can be simply expressed as:
\begin{equation}
{V}_{2}(x)=V_{0}(x)+2\eta^{'}(x)=V_{0}(x)-2\dfrac{d^{2}}{dx^{2}}\left(\ln W_{12}(x)\right),
\end{equation}
which will be free of singularities whenever $W_{12}(x)$ is nodeless. The eigenfunctions of $H_{0}$ and ${H}_{2}$, $\{\psi_{i}^{(0)}(x),$ ${\psi}_{i}^{(2)}(x), i=0, 1, \dots \}$, are connected by the operators $L_{2}^{-}$, $L_{2}^{+}$ as follows:

\begin{equation}
L_{2}^{-}\psi_{i}^ {(0)}=\dfrac{W(v_{1}, v_{2}, \psi_{i}^{(0) }(x))}{W_{12}(x)} \propto {\psi}_{i}^ {(2)}(x). \label{d1}
\end{equation}
From equation (\ref{d1}) it is clear that $L_{2}^{-}v_{1}(x)=L_{2}^{-}v_{2}(x)=0$. Moreover, the kernel of $L^{+}_{2}$ supplies two formal eigenfunctions of $H_{2}$ associated to $\epsilon_{1}$ and $\epsilon_{2}$:

\begin{equation}
{v}^{(2)}_{1}(x)\propto \dfrac{v_{2}(x)}{W_{12}(x)}, \quad {v}^{(2)}_{2}(x)\propto \dfrac{v_{1}(x)}{W_{12}(x)}. \label{marr1}
\end{equation}
Depending on whether they can be normalized or not, such factorization energies must be either included or not in the spectrum of $H_{2}$. Therefore, the Hamiltonians $H_{0}$ and $H_{2}$ are isospectral, with the possible exception of $\epsilon_{1}$ and $\epsilon_{2}$, which could be in Sp$(H_{2})$. From now on we will assume that $V_{0}(x)$ is a given initial periodic potential of period $T$, $V_{0}(x+T)=V_{0}(x)$. In this case there is no need to consider the range of the variable $x$ as the whole real line, but just a given period. According to the Bloch-Floquet theory, the physical wavefunctions $\psi(x)$ are quasi-periodic:

\begin{equation}
\psi(x+T)=e^{ikT}\psi (x) \label{joc1},
\end{equation}
where $k \in \mathbb{R}$ is called quasi-momentum or momentum of the crystal, and it defines a self-adjoint boundary value problem.

\subsection{Lam\'e equation: some remarks}

In order to study the effect of periodic potentials on bilayer graphene we start with the Lam\'e potential
\begin{equation}
V_{0}(x)=n(n+1)\textit{m}\,\text{sn}^ {2}(x|m). \label{n0}
\end{equation}
The stationary Schrödinger equation for the Hamiltonian (\ref{eo1}) with the periodic potential (\ref{n0}) is the Jacobi version of the \textit{Lam\'e equation}, where $\text{sn}(x|m)$ is the Jacobi elliptic function in Glaisher notation with modular parameter $m$. When $m \in (0, 1)$ the function $\text{sn}^{2}(x|m)$ has real period $2K=2K(m)$ and an imaginary period $2iK'=2iK(1-m)$, with $K(m)$ being the first complete elliptic integral. In addition, if $x$ is restricted to the real axis and $m$ and $n$ are real, the Lam\'e equation becomes a real domain Schrödinger equation with a periodic potential, i.e., a Hill's equation.

\medskip
The potential (\ref{n0}) has bounded physical solutions (\ref{joc1}) with an energy spectrum consisting of exactly $n+1$ allowed bands separated to each other by $n+1$ forbidden bands \cite{hill, maier}. The band edges arise when $kT=0, \pi, ...$, i.e., for
\begin{equation}
\psi(x+T)=e^{ikT}\psi(x)=\xi\psi(x),
\end{equation}
with $\xi=\pm 1$ being a Floquet multiplier. Due to the fact that $H_{0}$ in Eq. (\ref{eo1}) with $V_{0}(x)$ given by expression (\ref{n0}) is a Lam\'e operator of Hill type, we know from the oscillation theorem that there exists a sequence of real numbers (in general infinite)
\begin{equation}
E_{0}<E_{1}\leq E_{1'}<E_{2}\leq \dots <E_{j}\leq E_{j'}<\dots, \quad j=1, 2, ...
\end{equation} 
where the band edge energies $E_{j}$ and $E_{j'}$ correspond to $2K$-periodic solutions for $j$ even and $2K$-antiperiodic solutions for $j$ odd. The physical energies lie in allowed bands, which are intervals delimited by energies corresponding to the two values of $\xi$, i.e., to periodic and anti-periodic Bloch solutions. These allowed bands form a sequence, where $E_{0}$ is the first periodic eigenvalue if we come from $-\infty$, followed by alternating pairs of anti-periodic and periodic eigenvalues (each pair might be coincident). The allowed energy bands are $[E_{0}, E_{1}]$, $[E_{1'}, E_{2}], ..., [E_{j'}, E_{j+1}], ...$; consistently, the forbidden energy bands (energy gaps) are $(E_{1}, E_{1'})$, $(E_{2}, E_{2'}), ..., (E_{j}, E_{j'}), ...$ 
The band edge energies are usually called the discrete spectrum of the periodic potential. The corresponding eigenfunctions $\{\psi_{0}^{(0)}, \psi_{1}^{(0)}, \psi_{1'}^{(0)}, \psi_{2}^{(0)}, \psi_{2'}^{(0)}, \dots\}$ have periods $T$, $2T$, $2T$, $T$, $T$, $\dots$, and their respective number of nodes in a period $T$ is $0, 1, 1, 2, 2, \dots$. 

Many algebraic forms can be obtained from expression (\ref{n0}); nevertheless, for studying the properties of the Lam\'e equation it is important to take a form appropriate for the purposes at hand. For practical applications the Jacobian form, leading to the theta-functions, is the most suitable one. On the other hand, for studying the properties of the solutions it is better to use the second algebraic form, although in some problems the analysis is simpler using the Weierstrass form \cite{period2}. Let us stress that the Jacobi elliptic and Weierstrass functions have been widely used in the description of physical phenomena \cite{sch, pastras, lam1, Bri}. Recently, this interest has been renewed since they have been employed in direct methods for solving nonlinear differential equations \cite{ex1, ex2, ex3, ex4}.  

Let us express the Lam\'e equation in terms of the Weierstrass function $\wp=\wp (u; g_{2}, g_{3})$, which is a canonical elliptic function with a double pole at $u=0$ satisfying

\begin{equation}
(\wp')^ {2}=4(\wp-e_{1})(\wp-e_{2})(\wp-e_{3}).
\end{equation}
For ellipticity, the roots $\{e_{\gamma}\}_{\gamma=1}^{3}$ must be different, which is equivalent to ask that the modular discriminant $\Delta=g_{2}^{3}-27g_{3}^{2}$ should be non-zero. Either of $g_{2}, g_{3} \in \mathbb{C}$ may be equal to zero, but not both of them. The relation between the Jacobi and Weierstrass elliptic functions is well known. Choose, first $\{e_{\gamma}\}_{\gamma=1}^{3}$ in the way:
\begin{equation}
(e_{1}, e_{2}, e_{3})=A^ {2}\left(\dfrac{2-m}{3}, \dfrac{2m-1}{3}, -\dfrac{m+1}{3}\right),\label{n1}
\end{equation}
where $A \in \mathbb{C}/ \{0\}$ is any proportionality constant. Then, 
\begin{equation}
g_{2}=A^{4}\dfrac{4(m^{2}-m+1)}{3}, \qquad g_{3}=A^{6}\dfrac{4(m-2)(2m-1)(m+1)}{27},\label{n2}
\end{equation}
and the dimensionless ($A$-independent) Klein invariant $J=g_{2}^{3}/\Delta$ is given by
\begin{equation}
J=\dfrac{4}{27}\dfrac{(m^{2}-m+1)^{3}}{m^ {2}(1-m)^{2}}.
\end{equation}
Two sorts of elliptic function are related with $\wp$ as follows:
\begin{equation}
\text{sn}^{2}(Az|m)=\dfrac{e_{1}-e_{2}}{\wp (z)-e_{3}}, \qquad \text{ns}^{2}(Az|m)=\dfrac{\wp (z)-e_{3}}{e_{1}-e_{3}}, \label{nn2}
\end{equation}
thus the periods of $\wp$, denoted $2 \omega$, $2 \omega '$, will be related to those of $\text{sn}^{2}$ by
\begin{equation}
2\omega=\dfrac{2K}{A}, \qquad 2\omega '=\dfrac{2iK'}{A}.
\end{equation}
The case when $2K$, $2K'$ are real, or equivalently $\omega \in \mathbb{R}$, $\omega ' \in i\mathbb{R}$, corresponds to the case when $g_{2}, g_{3} \in \mathbb{R}$ and $\Delta >0$. Choosing now for simplicity $A=1$, so that $e_{1}-e_{3}=A^{2}=1$, we can rewrite the Lam\'e equation with the aid of (\ref{nn2}) in its Weierstrass form:
\begin{equation}
\left\{\dfrac{d^{2}}{du^ {2}}-[n(n+1)\wp (u; g_{2}, g_{3})+B]\right\}\psi=0,
\end{equation}
where $u=x+iK'$. Note that the translation of (\ref{n0}) by $iK'$ replaces $m {\,}sn^{2}$ by  $\text{ns}^{2}$. Moreover, $B=-E(e_{1}-e_{3})-n(n+1)e_{3}$, \textit{i.e.}, 
\begin{equation}
B=-E+\dfrac{1}{3}n(n+1)(m+1),
\end{equation}
is a transformed energy parameter. 

Because for $n=1$ the hyperbolic limit yields a bound state at $E=0$, and scattering states above $E=1$, for the periodic case we would expect to see at least an energy band with one edge at $E=0$. Therefore, another band edge energies ought to be below $E=1$. The band edge eigenstates found explicitly are $\text{dn}(x, m)$, $\text{cn}(x, m)$ and $\text{sn}(x, m)$, with an allowed band energy going from $E=0$ to $E=1-m$, then a band gap, and finally an infinite band energy starting from $E=1$. In order to find the remaining states, a formal solution of the eigenvalue problem for the Hamiltonian $H_{0}$ is necessary. This problem can be expressed as
\begin{equation}
\psi ''(x)=(2m{\,}\text{sn}(x, m)^ {2}-m-E)\psi(x), \label{lup1}
\end{equation}
or, alternatively, 
\begin{equation}
\dfrac{d^{2}\psi}{du^ {2}}-[2\wp (u; g_{2}, g_{3})+B]\psi=0,
\end{equation}
which is a special case of the Lam\'e differential equation, whose solution is shown in \cite{period2, period1, period3}. In the considered case $(n=1)$ the solution is given by the following independent functions 
\begin{equation}
\psi(x, k)=\dfrac{\sigma (x+k)}{\sigma(x)\sigma (k)}\exp(-\zeta (k)x),
\end{equation}
where $k\in \mathbb{C}$ and $\sigma (x)$ and $\zeta (x)$ are the Weierstrass $\sigma$ and $\zeta$ functions. Such solutions have the Floquet property 
\begin{equation}
\psi(x+2\omega, k)=\exp(2\eta k-2\sigma (k)\omega) \psi (x, k),
\end{equation} 
with $\eta=\zeta(\omega)$. In terms of Jacobi elliptic functions the potential (\ref{n0}) for $n=1$ shifted by the energy $-m$, has the following band edge eigenvalues and eigenfunctions:
\begin{equation}
E_{0}=0, \qquad \psi_{0}^{(0)}(x)=\text{dn} (x), \label{miles1}
\end{equation}
\begin{equation}
E_{1}=1-m, \qquad \psi_{1}^{(0)}(x)=\text{cn} (x), \label{miles2}
\end{equation}
\begin{equation}
E_{1'}=1, \qquad \psi_{1'}^{(0)}(x)=\text{sn} (x). \label{miles3} 
\end{equation}

\section{Bilayer graphene in periodic magnetic fields}

For a periodic potential $V_{0}(x)$ every finite energy gap is limited by two eigenfunctions $\psi_{j}^ {(0)}(x)$ and $\psi_{j'}^{(0)}(x)$, $j=1, 2, ...,$ which have the same number of nodes ($j$). This suggests to take both eigenfunctions to implement a second-order SUSY transformation because we will obtain a physically acceptable partner Hamiltonian $H_{2}$ with a non-singular, periodic potential $V_{2}(x)$. Due to the periodicity $T$ of the intertwining operator $L_{2}^{-}$ any eigenfunction $\psi_{i}^{(0)}$ of the discrete spectrum of $H_{0}$ will be transformed by the action of $L_{2}^{-}$ into an eigenfunction $\psi_{i}^{(2)}(x)$ of the discrete spectrum of $H_{2}$. Thus, $H_{0}$ and ${H}_{2}$ are isospectral Hamiltonians sharing the same band structure \cite{dar0}.

\medskip

If we replace now the band edge solutions $\psi_{1}^{(0)}$ and $\psi_{1'}^{(0)}$ in equations (\ref{magpot}-\ref{q2}) it turns out that the two SUSY partner potentials differ just by a displacement in the argument i.e., $V_{2}(x)=V_{0}(x+K(m))$. The form of the potentials is formally different, but in fact there is no any new physical information due to 

\begin{equation}
V_{0}(x)=2m{\,}\text{sn}^ {2}(x, k)-m, \label{or1}
\end{equation}
\begin{equation}
V_{2}(x)=2m{\,}\text{sn}^ {2}(x+K(m), k)-m. \label{or2}
\end{equation}
This happens since it is known that a self-isospectral condition is fulfilled if $\psi_{j}^{(0)}(x)\psi_{j}^{(0)}(x+T/2) \propto \psi_{j'}^{(0)}(x)\psi_{j'}^{(0)}(x+T/2)$ \cite{dar0, dar1}. In our case, choosing the band edge eigenfunctions $\psi_{1}^{(0)}$ and $\psi_{1'}^{(0)}$ given in equations (\ref{miles2}) and (\ref{miles3}) we will have
\begin{equation}
\text{cn}(x, m)\text{cn}(x+T/2, m) \propto \text{sn}(x, m) \text{sn}(x+T/2, m),
\end{equation}
which is the condition for $V_{2}(x)$ to be physically equivalent (self-isospectral) to $V_{0}(x)$. Thus, there is no new potential obtained for $n=1$ if band edge eigenfunctions are used. However, for the Lam\'e potential (\ref{n0}) with $n=2, 3, ...$, the self-isospectral condition is not longer fulfilled, and thus new potentials could be obtained \cite{dunne,  kha, naga, dunne2}. 

\medskip
Let us point out that the general Darboux-Crum transformations have been applied to the single gap periodic Lam\'e potential to generate an arbitrary countable set of bound states in its two forbidden bands, the infinite lowest and the finite intermediate one. As a consequence, nonperiodic single gap potentials arise, which contain two essentially different types of soliton defects in the periodic background \cite{dar2, exo1, exo2}. As it was already shown, these defects appear in a natural way in carbon-based crystals such as monolayer graphene \cite{cel1} and, as we will see below, for bilayer graphene. 
In order to do that, we need to use as seed solutions linear combinations of the two generalized Bloch functions belonging to the same spectral gap in order to obtain the new potential.  In this way, the second-order transformation will produce a quasi-periodic magnetic superlattice supporting two bound states.

\begin{figure}[ht]
\centering
\begin{subfigure}{0.45\textwidth}
    \includegraphics[width=\textwidth]{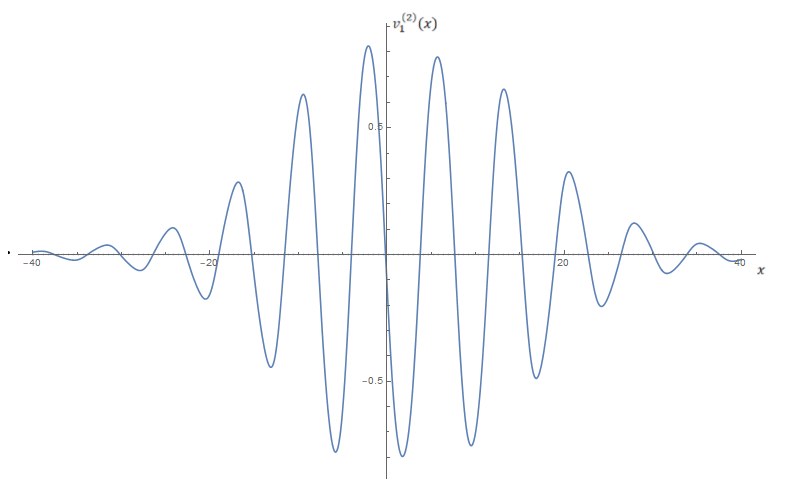}
    \caption{}
    \label{fig:m1}
\end{subfigure}
\begin{subfigure}{0.45\textwidth}
    \includegraphics[width=\textwidth]{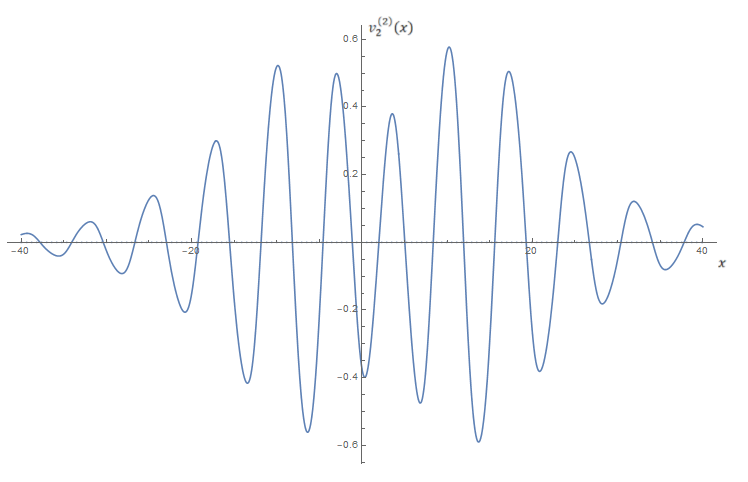}
    \caption{}
    \label{fig:m2}
\end{subfigure}    
\caption{Bound states created at the energy $\alpha_{1}=0.9$ (a) and at $\alpha_{2}=0.8$ (b) for $m=0.5$.} \label{dani2}  
\end{figure}

Let us start from the Lam\'e magnetic superlattice; the seed solutions we will consider, which fulfill the Lam\'e equation, are $v_{j}=u_{1}(x, \epsilon_{j})+\eta_{j}u_{2}(x, \epsilon_{j})$, $j=1, 2$ with $\epsilon_{2}<\epsilon_{1}<0$. The two Bloch functions $u_{j}$ take the form

\begin{equation}
u_{1}(x, \epsilon_{j})\!\!=\!\!\dfrac{\sigma(x_{0}+\omega ')\sigma(x+\delta_{j}+\omega ')}{\sigma(x+\omega ')\sigma(x_{0}+\delta_{j}+\omega ')}e^{-\zeta (\delta_{j})(x-x_{0})}, \,\, u_{2}(x, \epsilon_{j})\!\!=\!\!\dfrac{\sigma(x_{0}+\omega ')\sigma(x-\delta_{j}+\omega ')}{\sigma(x+\omega ')\sigma(x_{0}-\delta_{j}+\omega ')}e^{\zeta (\delta_{j})(x-x_{0})}, \label{ma1}
\end{equation}
where $j=1, 2$, $x_{0}$ is a fixed point in $[0, T=2K]$, $\omega=K$ and $\omega'=iK'$ are the real and imaginary half-periods of the Jacobi elliptic functions, $\sigma$ and $\zeta$ are the nonelliptic Weierstrass functions, while the factorization energy $\epsilon_{j}$ and the complex displacement $\delta_{j}$ are related by
\begin{equation}
\epsilon_{j}=\dfrac{2}{3}(m+1)-\wp (\delta_{j})-m.
\end{equation}
A short hand notation for $u(x, \epsilon_{j})$ in the Wronskians will be employed from now on, 
\begin{equation}
W(v_{1}, v_{2})=W(\delta_{1}, \delta_{2})+\eta_{1}W(-\delta_{1}, \delta_{2})-\eta_{2}W(\delta_{1}, -\delta_{2})-\eta_{1}\eta_{2}W(-\delta_{1}, -\delta_{2}),\label{om0}
\end{equation}
with
\begin{equation}
W(\delta_{1}, \delta_{2})=\dfrac{\sigma^{2}(x_{0}+\omega ')\sigma (\delta_{2}-\delta_{1})}{\sigma (\delta_{1})\sigma (\delta_{2})\sigma (x_{0}+\delta_{1}+\omega ')\sigma (x_{0}+\delta_{2}+\omega ')} \dfrac{\sigma (x+\delta_{1}+\delta_{2}+\omega ')}{\sigma (x+\omega ')}e ^{-[\zeta (\delta_{1})+\zeta (\delta_{2})](x-x_{0})}, \label{om1}
\end{equation}
where we have used the symmetry of the quasi-periodic elliptic sigma and zeta functions $\sigma(-x; g_{2}, g_{3})=-\sigma(x; g_{2}, g_{3})$ and $\zeta(-x; g_{2}, g_{3} )=-\zeta(x; g_{2}, g_{3})$, respectively. There exist an infinite closed sector in $\mathbb{R} ^{2}$, bounded by the $\eta_{1}-$ and $\eta_{2}-$ axes, where the general Wronskian (\ref{om0}) is nodeless. Specifically, the Wronskian defined by equation (\ref{om1}) keeps constant sign and it is strictly positive for $\eta_{1}, \eta_{2} \geq 0$, enabling a nonsingular Darboux transformation.  In order to obtain the explicit expression for the magnetic field generated by supersymmetry, we replace the expressions (\ref{om0}-\ref{om1}) in $(\ref{magpot})$. 
In particular, for the limit case $\eta_{1} \to 0$ and $\eta_{2}\to 0$ the Wronskian (\ref{om0}) becomes (\ref{om1}), which is consistent with \cite{dar1}. In the general case, the explicit expression for the magnetic field with impurities is obtained by replacing Eq. (\ref{ma1}) in (\ref{kim2}): 

\begin{align}
    B(x; \epsilon_{1}, \epsilon_{2}) & \propto dn^{2}(x|m)-dn^ {2}(x+\delta_{1}+\delta_{2}|m)+\eta_{1}\eta_{2}(dn^{2}(x-\delta_{1}-\delta_{2}|m)-dn^{2}(x|m))\nonumber \\
         & +\eta_{1}(dn^ {2}(x|m)-dn^ {2}(x-\delta_{1}+\delta_{2}|m))+\eta_{2}(dn^ {2}(x|m)-dn^ {2}(x+\delta_{1}-\delta_{2}|m)), \label{qmag1}
\end{align}
with asymptotic behaviour for $x\to\pm \infty$ given by 
\begin{equation}
B(x \to \infty; \epsilon_{1}, \epsilon_{2}) \to  \eta_{1}\eta_{2}(dn^{2}(x-\delta_{1}-\delta_{2}|m)-dn^{2}(x|m)), \label{qmag2}
\end{equation}
\begin{equation}
B(x \to -\infty; \epsilon_{1}, \epsilon_{2}) \to  dn^{2}(x|m)-dn^ {2}(x+\delta_{1}+\delta_{2}|m). \label{qmag3}
\end{equation}
Expression (\ref{qmag1}) represents a Lamé magnetic superlattice with two additional impurities, analogous to the case obtained for monolayer graphene with a single displacement and an added impurity \cite{cel1}. It is straightforward to find the expressions for the added bound states in the forbidden band $(E_{1},E_{1'})$ for bilayer graphene, which arise from equation (\ref{marr1}) using the linear combinations $v_{1}$ and $v_{2}$ of the Bloch functions (see a plot in Fig. \ref{dani2}). Also, we can generate the partner potentials $V_{0}(x)$ and $V_{2}(x)$ given by expressions (\ref{q1}) and (\ref{q2}) for our linear combination of Bloch functions, once the Wronskian (\ref{om1}) is found for the arbitrary displacements $\delta_{j}$. Let us stress that the implementation of the second order supersymmetry transformation for bilayer graphene is mandatory, unlike what happens for monolayer graphene where an additional first order SUSY transformation could be applied to go deeper in the study of impurity effects, but it is not strictly required.  

\begin{figure}[ht]
\includegraphics[width=0.5\textwidth]{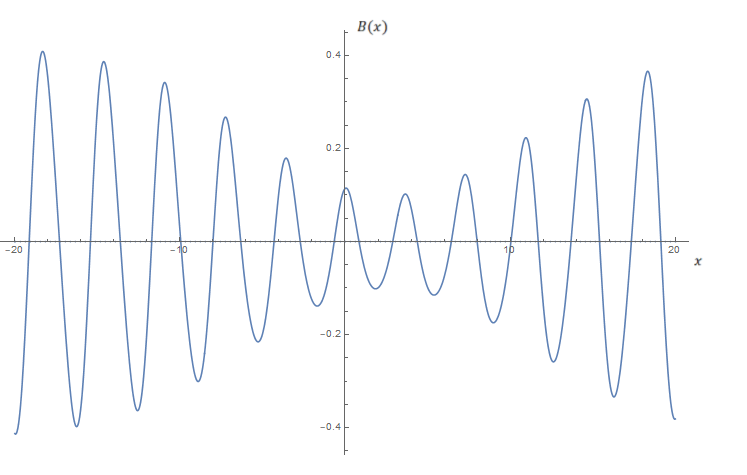}
\centering
\caption{Quasiperiodic magnetic superlattice generated by applying a second order SUSY transformation to bilayer graphene.}\label{eros0}
\end{figure}

The magnetic field amplitude $(\ref{qmag1})$ is depicted in Figure \ref{eros0}. From such explicit expression, and the corresponding asymptotic behaviours (\ref{qmag2}) and (\ref{qmag3}) for $x\pm \infty$, we can see that the magnetic field recovers its periodicity asymptotically. 
Throughout this work we have restricted ourselves to study the effect of the Lam\'e periodic potentials in bilayer graphene; nevertheless, the basic ideas of this work could be applied to more general potentials, e.g., to the associated Lam\'e potentials \cite{fer0, fer0001}. We are confident that the basic idea of introducing defects in the periodic magnetic superlattice for bilayer graphene, by adding bound states in the forbidden bands, can be well understood through the simplest periodic potentials at hand. 
\section{Conclusions}
Van der Waals heterostructures, such as bilayer graphene, offer an alternative amidst the advent and rapid progress of nanotechnology. Their development has increased the availability of atomic-layer materials, encompassing semiconductors, semimetals, superconductors, ferromagnets, and topological insulators. This progress has parallel advancements in theoretical frameworks, facilitating the rapid creation of quantum metamaterials for electrons. The stacking of two-dimensional layers atop one another forms a superlattice, where electrons can be described by Bloch bands within an enlarged unit cell. These lattices feature a corresponding tunable Brillouin zone, enabling precise control over the arrangement of electron and hole filling in the heterostructure. The ability to engineer unit cell sizes and electron filling can yield significant outcomes, particularly when applying a magnetic field. Quantum metamaterials aim to manipulate and control bound states embedded in superlattices and electromagnetic waves at the quantum level, enabling functionalities such as tunable optical properties, efficient light-matter interactions and novel quantum devices like quantum sensors or transistors.
In this direction, starting from the tight binding model for the low-lying excitations of bilayer graphene, we have determined the energy band structure of bilayer graphene when a periodic magnetic field is applied. In this approach, which generalizes previous works about monolayer graphene, we have employed second order SUSY transformations to deal with the topological features of electronic states in bilayer graphene. The bound states created by the transformations are often related to impurities or van Hove singularities. Alike to the case of monolayer graphene, the implementation of the most general Darboux transformation breaks the periodicity of the originally periodic magnetic field, due to the introduction of such bound states. However, afar from these introduced defects in graphene the solutions turn out to be periodic, behaving as it was expected. It is worth to note that the number of experimental studies about electronic transitions have increased in the last years, with its optimization being a major concern. About this point, the present study mimics the effect of gold island-enhanced multiplex quantum dots, in which the quantum dots act as a bridge between the gold islands. On the other hand, the present analysis reinforces the role of the second order supersymmetric quantum mechanics for studying electronic properties of bilayer graphene. Throughout this work we restricted ourselves to the simplest case of the Lam\'e potentials; nevertheless, we might choose, between a long list of candidates, another examples to illustrate the effects of the Darboux transformations for bilayer graphene.

\section*{Acknowledgments}
The authors acknowledge the support of Consejo Nacional de Humanidades Ciencia y Tecnolog\'{\i}a (CONAHCyT-M\'exico) under grant FORDECYT-PRONACES/61533/2020. OPT acknowledges CONAHCyT by a postdoctoral fellowship.

\medskip

\noindent \textbf{Data Availability Statement} Data sharing not applicable to this article as no datasets were generated or analyzed during the current study.

\end{document}